# Regulation Theory

*F. Bouvet*
Synchrotron SOLEIL, Gif-sur-Yvette, France

**Abstract**
This paper reviews the design of regulation loops for power converters. Power converter control being a vast domain, it does not aim to be exhaustive. The objective is to give a rapid overview of the main synthesis methods in both continuous- and discrete-time domains.

## 1     Introduction

Power converters convert electric energy from one form to another which is optimally tailored for user loads. Therefore regulation is an important part of the design and construction of any power converter.

Owing to the fast development of digital electronics, power converter control uses more and more embedded computing systems like microcontrollers or Digital Signal Processors (DSPs). These devices replace progressively analogue controllers that have dominated the control of power electronics systems for many years.

After a period during which the discrete-time control laws were directly obtained from a discretization of the continuous-time control laws in order to ensure continuity with the existing methods and equipment, automation has refined its tools and methodologies: new approaches based on discrete-time models of the processes to be controlled have been proposed. With these new techniques, more powerful synthesis methods have been developed. Alternatives to the still widely used PID-type controllers have emerged, like RST controllers or more exotic ones such as dead-beat or fuzzy controllers. All these alternatives offer potential improvements in performance.

In the first part of this paper, the main concepts of continuous-time control are reviewed. Only the control of single-input single-output linear and time-invariant systems is addressed here. Digital control is presented in the second part. After a few reminders about Z-transforms, the concept of the discrete-time model is introduced. The main digital controller synthesis methods are then described and, depending on the chosen method, the choice of the sampling frequency is discussed.

## 2     Continuous-time control

### 2.1     Memory refreshing

To design a controller that makes a system behave in a specific desirable manner, we need a way to predict its behaviour over time, specifically how its outputs will change in response to any applied inputs. Thus a mathematical description ('model') of the system to be controlled (also called the 'plant') is needed.

If the plant is a single-input single-output linear and time-invariant system (time-invariant systems are systems whose characteristics do not change with time), then its input $u(t)$ and output $y(t)$ are related by the following differential equation with constant coefficients:

$$a_n \cdot \frac{d^n}{dt^n} y + \ldots + a_1 \cdot \frac{d}{dt} y + a_0 \cdot y = b_m \cdot \frac{d^m}{dt^m} u + \ldots + b_1 \cdot \frac{d}{dt} u + b_0 \cdot u \quad (m \leq n), \qquad (1)$$

$n$ being the order of the system.

In the frequency domain, transfer functions are obtained from time-domain descriptions via Laplace transform. Assuming zero initial conditions, Eq. (1) leads to the following *system transfer function* $H(s)$:

$$H(s) = \frac{Y(s)}{U(s)} = \frac{\sum_{i=0}^{m} b_i \cdot s^i}{\sum_{j=0}^{n} a_j \cdot s^j}, \qquad (2)$$

where $s$ is the Laplace operator.

Some basic but important definitions are recalled:

- The *poles* are the roots of the denominator polynomial of the system transfer function $H(s)$. The *zeros* are the roots of the numerator polynomial.
- The system stability is determined by the location of the poles. If their real part is strictly negative, then the system is stable.
- If the transfer function behaves like $K/s^\alpha$ for $s \to 0$, then $\alpha$ is called the *system class*. The static gain is obtained for $s$ equal to zero.
- If there is a period of time during which the output does not react to the input, the system is defined as a system with delay. The transfer function $H_d(s)$ of a system having a time delay of $t0$ can be expressed as

$$H_d(s) = H(s) \cdot e^{-s \cdot t0} \qquad (3)$$

where $H(s)$ is the transfer function of the system without delay.

## 2.2 Importance of feedback control

Feedback is a very powerful mechanism. Why is it necessary? An intuitive strategy would be to use a feedforward controller and to make the product of the plant and this feedforward controller unity, hence to find a controller that equals the plant inverse ($C = H^{-1}$ in Fig. 1, assuming a unit gain for the actuator transfer function). Then in theory the system output $Y$ would be equal to the controller input $Yref$.

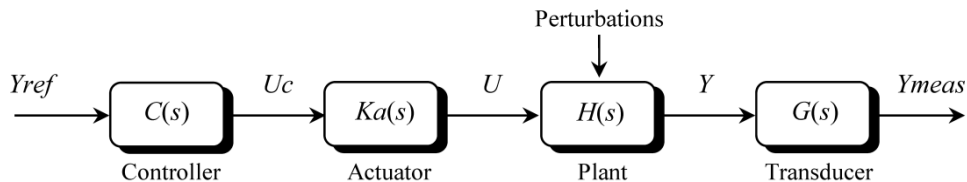

**Fig. 1:** Open-loop control configuration

This works well as long as the feedforward controller approximates the plant inverse sufficiently well. In practice, uncertainties and unmodelled dynamics make this impossible. To fulfil the requirements, a feedback correction is necessary to guarantee performance, even with model uncertainties, to reduce sensitivity to parameter variations, to reject disturbances, and obviously to stabilize unstable open-loop systems. Thus a typical control structure combines feedback and feedforward controls (see Fig. 2). The feedback controller guards robust stability and improves static and dynamic precision whereas the feedforward controller improves tracking behaviour (via compensation of the input) or regulation behaviour (via compensation of the perturbations).

It is important to understand that the performance requirements (static, dynamic behaviour, stability, and robustness) of the closed loop translate into constraints on the compensated system in open loop. We will see later in this paper how to make use of the open-loop frequency response to design controllers.

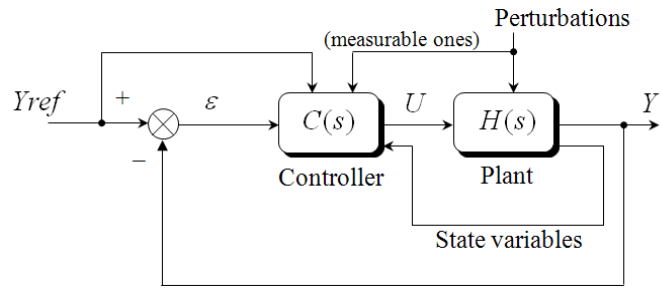

**Fig. 2:** Typical control structure

The typical control structure for current regulated power supplies is represented in Fig. 3. This so-called 'cascade' structure is based on a nested connection of a fast inner voltage loop and a slower outer current loop. The fast inner voltage loop acts as an active filter to reject the output voltage ripple and the output voltage fluctuations due to float of input mains, and it also simplifies the design of the current loop. The outer current loop ensures the overall stability of the power supply and it provides an inherent overvoltage protection.

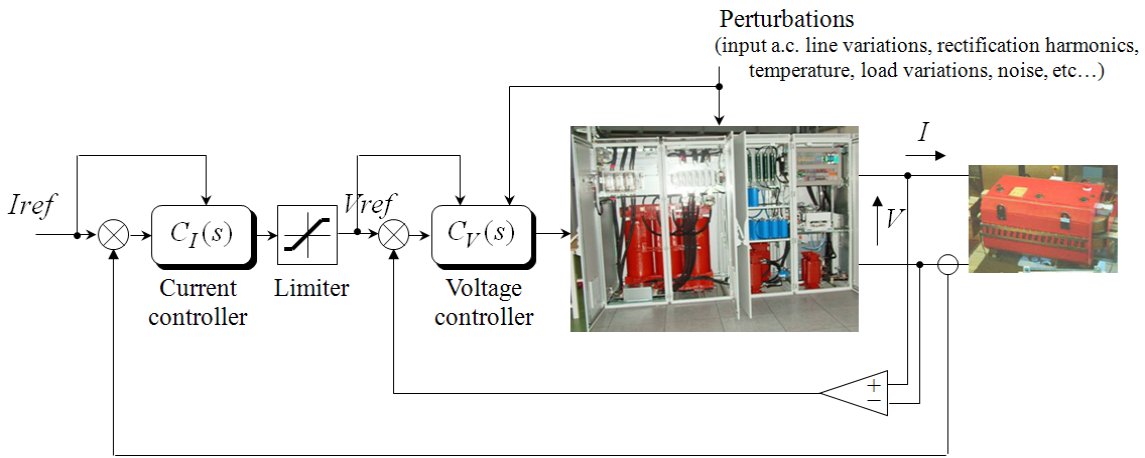

**Fig. 3:** Cascade control structure

## 2.3 Analysis of closed-loop systems

### 2.3.1 *Definition of open-loop and closed-loop transfer functions*

In this section, we focus on the closed-loop system shown in Fig. 4. To study it, we use the open-loop transfer function obtained by breaking the feedback loop, the expression for which is given by

$$OL(s) = \frac{Ymeas(s)}{\varepsilon(s)} = C(s) \cdot H(s) \cdot G(s). \qquad (4)$$

The open-loop bandwidth is defined as the interval of angular frequencies for which the open-loop gain $|C(jw) \cdot H(jw) \cdot G(jw)|$ is higher than unity.

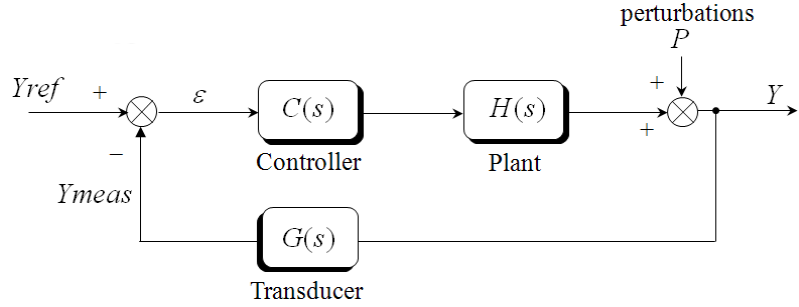

**Fig. 4:** Closed-loop system

The closed-loop transfer function is

$$CL(s) = \frac{Y(s)}{Yref(s)} = \frac{C(s) \cdot H(s)}{1 + C(s) \cdot H(s) \cdot G(s)}. \qquad (5)$$

The transfer functions of the error versus input and perturbation can be calculated as

$$\frac{\varepsilon_{Yref}(s)}{Yref(s)} = \frac{1}{1 + C(s) \cdot H(s) \cdot G(s)}$$
$$\frac{\varepsilon_P(s)}{P(s)} = -\frac{G(s)}{1 + C(s) \cdot H(s) \cdot G(s)} \qquad \varepsilon(s) = \varepsilon_{Yref}(s) + \varepsilon_P(s). \qquad (6)$$

### 2.3.2 Precision of closed-loop systems

#### 2.3.2.1 Static error

In steady state, the error versus the input can be derived from Eq. (6). Using the 'final value' theorem, we have

$$\lim_{t \to \infty} \varepsilon_{Yref}(t) = \lim_{s \to 0} s \cdot \varepsilon_{Yref}(s) = \lim_{s \to 0} \frac{s}{1 + C(s) \cdot H(s) \cdot G(s)} \cdot Yref(s). \qquad (7)$$

- For a step input ($Yref(s) = K/s$), the error cancels if there is at least one integrator in the open-loop transfer function.
- For a ramp input ($Yref(s) = K/s^2$), two integrators are needed to achieve zero steady-state error.
- For a sinusoidal input $K \cdot \sin(w_0 \cdot t)$, at steady state, the error is a harmonic signal which module $|\varepsilon_{Yref}|$ is such that $\dfrac{|\varepsilon_{Yref}|}{K} = \left|\dfrac{1}{1 + C(s) \cdot H(s) \cdot G(s)}\right|_{s=jw_0}$.

So if $w_0$ is inside the open-loop bandwidth, the error amplitude is inversely proportional to the open-loop gain at $w_0$: $\dfrac{|\varepsilon_{Yref}|}{K} \approx \left|\dfrac{1}{C(s) \cdot H(s) \cdot G(s)}\right|_{s=jw_0}$.

Regarding perturbation rejection, using Eq. (6), we have

$$\lim_{t \to \infty} \varepsilon_P(t) = \lim_{s \to 0} s \cdot \varepsilon_P(s) = \lim_{s \to 0} \frac{-s \cdot G(s)}{1 + C(s) \cdot H(s) \cdot G(s)} \cdot P(s). \quad (8)$$

To have a perfect rejection, the open loop must contain the classes of the perturbations. Thus to reject disturbances of class *N*, at least *N* integrators are needed in the open-loop transfer function.

*2.3.2.2  Dynamic error*

Let us assume that the velocity *v* and the acceleration *γ* of the input are limited. The input signal is then defined by the following constraints: $v < v_{max}$, $\gamma < \gamma_{max}$.

Maintaining the dynamic error $\varepsilon_d$ below a given limit $\varepsilon_{d_{max}}$ results in a specification of a minimum open-loop gain in a certain frequency range:

$$\varepsilon_d < \varepsilon_{d_{max}} \quad \Rightarrow \quad |OL(s)|_{s=j\frac{\gamma_{max}}{v_{max}}} > \frac{v_{max}^2}{\gamma_{max} \cdot \varepsilon_{d_{max}}}. \quad (9)$$

In general, this constraint results in imposing a finite open-loop bandwidth.

*2.3.3  Stability and robustness of closed-loop systems*

A closed-loop system is stable if the poles of its transfer function have a negative real part.

The stability of a feedback system can also be determined through the simple knowledge of the open-loop frequency response, using the 'Nyquist criterion'. This is of great practical importance for controller design as it easily demonstrates how to modify the controller to make an unstable system stable. The closed-loop system is stable if the critical point −1 is on the left-hand side of the open-loop Nyquist curve, for *w* increasing. Two quantitative measures are commonly used to determine how stable (robust) the system is. The *phase margin* is the amount of phase lag required to reach the stability limit. The *gain margin*, similarly, states how much the controller gain can be increased before reaching the stability limit.

**Phase margin:**

$$\Phi_M = 180° + \arg\left[OL(j \cdot w_{cr})\right], \quad (10)$$

where $w_{cr}$ is such that $|OL(j \cdot w_{cr})| = 1$.

**Gain margin:** $\quad G_M = \dfrac{1}{|OL(j \cdot w_\pi)|}, \quad (11)$

where $w_\pi$ is such that $\arg\left[OL(j \cdot w_\pi)\right] = -180°$.

Typically, $\quad 30° < \Phi_M < 60°$, $\quad G_M > 6$ dB.

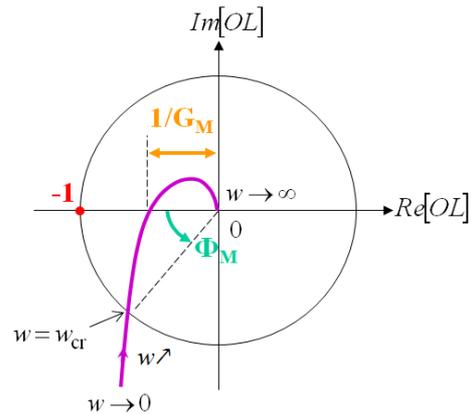

**Fig. 5:** Closed-loop system stability

There is an obvious contradiction between precision and stability specifications. Thus the controller design will be the result of a trade-off.

## 2.3.4 Influence of the poles on the transient behaviour

The contribution of real and complex poles on the transient behaviour of a system is represented in the complex plane in Fig. 6. An important observation is that the most distant poles to the left yield a faster transient regime. So the poles closest to the imaginary axis are the ones that tend to dominate the response since their contribution takes a longer time to die out. These poles are called *dominant poles* if the ratio of their real part to that of any other poles is typically lower than 1/5. A transfer function can be simplified by keeping only the dominant poles (and the static gain unchanged). This makes control loop analysis and design easier. Also, lower-order controllers can be obtained.

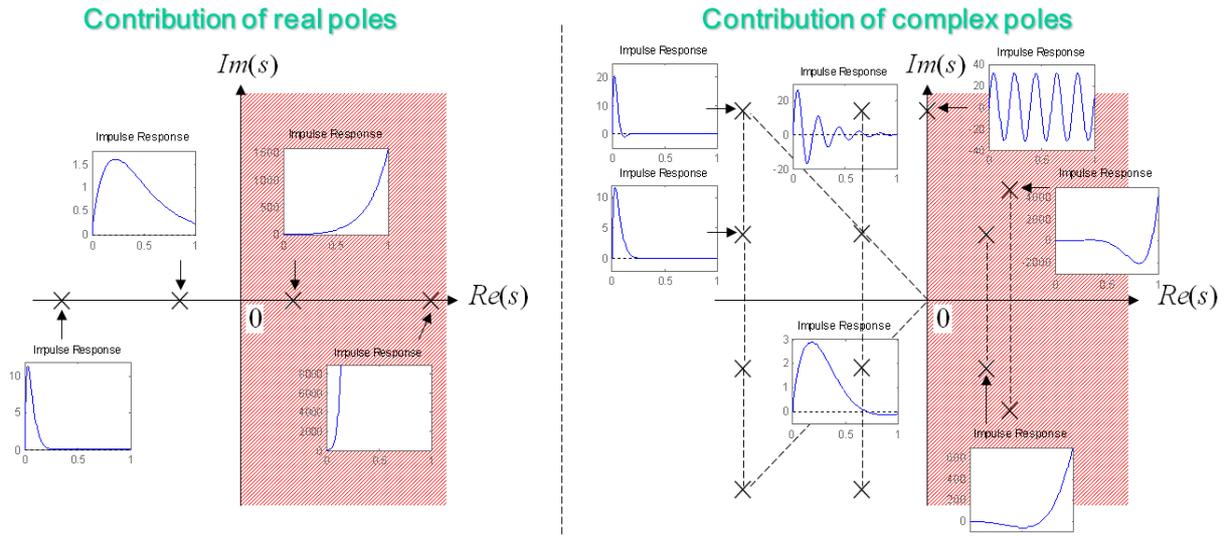

**Fig. 6:** Influence of the poles on the system transient behaviour

## 2.3.5 Particular case: second-order systems with complex conjugate poles

The search for a good compromise between speed and stability generally leads to the choice of a first- or second-order behaviour for the closed loop. For second-order system behaviour, the design specifications imply constraints on the *cut-off frequency* $w_n$ and the *damping ratio* $\zeta$ of the targeted transfer function Eq. (12). Figure 7 shows the parameters usually used to characterize the step response, and Eqs. (13)–(15) link cut-off frequency and damping ratio to these parameters.

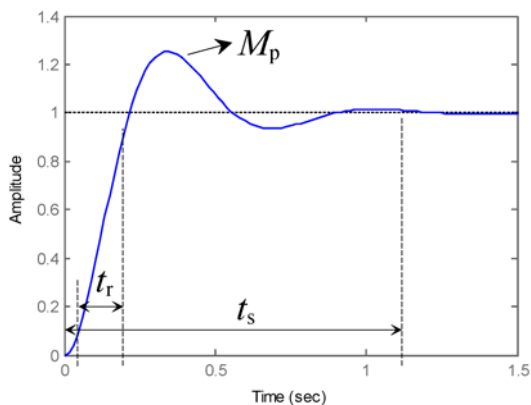

**Fig. 7:** Step input response

Targeted closed-loop transfer function:

$$CL_{\text{des}}(s) = \frac{w_n^2}{s^2 + 2 \cdot \zeta \cdot w_n \cdot s + w_n^2} \; ; \qquad (12)$$

Rise time (10% → 90%):

$$t_r \approx (2.6 \cdot \zeta^2 - 0.45 \cdot \zeta + 1.2)/w_n \; ; \qquad (13)$$

Peak overshoot: $\quad M_p = e^{-\pi \cdot \zeta / \sqrt{1-\zeta^2}} \; ; \qquad (14)$

Settling time (to 1%): $\quad t_s \approx 4.6/\zeta \cdot w_n \; . \qquad (15)$

## 2.3.6 Influence of a zero on the system behaviour

In the case of a second-order system, the corresponding transfer function is

$$CL(s) = K \cdot (s + z_0) / (s - p_1) \cdot (s - \overline{p_1}), \quad (16)$$

where $z_0$ is assumed to be a strictly negative real number.

So the unit step response of this system can be written as

$$Y(s) = \frac{K}{s} \cdot \frac{s + z_0}{(s - p_1) \cdot (s - \overline{p_1})} = \frac{K \cdot z_0}{s \cdot (s - p_1) \cdot (s - \overline{p_1})} + \frac{K}{(s - p_1) \cdot (s - \overline{p_1})}, \quad (17)$$

which gives in the time domain

$$y(t) = y_{\text{2nd order}}(t) + \frac{1}{z_0} \cdot \frac{d}{dt} y_{\text{2nd order}}(t), \quad (18)$$

where $y_{\text{2nd order}}(t)$ is the step response of the system Eq. (16) without zeros.

The second term of Eq. (18) reveals that the additional zero makes the system faster and more oscillatory, and all the more so as its value is close to zero. So special care should be taken to design appropriate zeros in a closed-loop transfer function.

### 2.4 Continuous-time controller synthesis

#### 2.4.1 Controller design process

In the design of a controller, the first step is to get the dynamic model of the system to be controlled. The principal difficulty in modelling power converters is that they are inherently non-linear and present several distinct electric configurations during a switching period. By constructing equivalent averaged circuit models, large-signal average models of the power converter can be determined. Then linearization about a quiescent operating point may be necessary to obtain linear small-signal transfer functions [1–4]. To elaborate power converter models, an identification processes from experimental data can also be used.

Once the model is known, the specifications of the desired closed-loop performance have to be defined. As explained in the preceding section, the choice of these specifications results from a trade-off between speed and robustness. This choice is obviously linked to the plant dynamics but also to the power availability of the power converter during the transient: the acceleration of the plant natural response requires control peaks that are greater than the steady-state values. If the control variable comes to saturation, the feedback loop is broken and the actuator remains at its limit independently of the plant output. To prevent actuator saturation, the closed-loop performance has to be chosen accordingly.

The last step is to choose the controller type and its design method.

#### 2.4.2 Proportional-integral-derivative (PID) controller

##### 2.4.2.1 Introducing PID control

The PID controller is by far the most dominant form of feedback in use today [5–7], especially for analogue control. PID feedback is simple and intuitive: it involves only three separate constant parameters to tune the control loop. As shown in Eqs. (19) and (20), it is based on the past control error (integral term), the present control error (proportional term), and the future control error (derivative term). Many controllers (PI) do not even use derivative action. This kind of controller is well suited for systems exhibiting dominant first- or second-order behaviour, for which the desired performance of the closed-loop as compared to the open-loop response of the system is not too demanding.

PID algorithm:
$$u(t) = K_p \cdot \varepsilon(t) + K_i \int_0^t \varepsilon(t) + K_d \cdot \frac{d}{dt}\varepsilon(t); \qquad (19)$$

Controller transfer function:
$$C_{PID}(s) = K_p + \frac{K_i}{s} + \frac{K_d \cdot s}{1 + \frac{K_d}{N \cdot K_p} \cdot s}, \quad 10 \leq N_{typ.} \leq 20. \qquad (20)$$

As a pure derivative amplifies noise, the transfer function standard form includes a low-pass filter on the derivative term.

A number of alternative approaches for PID tuning are available, including the following ones:

– heuristic tuning procedures: Ziegler–Nichols, Cohen–Coon, etc.,
– graphical methods: loop shaping, root locus, etc.,
– pole placement,
– minimization of integral type criterion.

*2.4.2.2  Putting it into practice*

Let us focus on two design methods—loop shaping and pole placement—in the following example where controllers have to be designed for the current and voltage control loops of a Buck converter feeding an inductive (R-L) load.

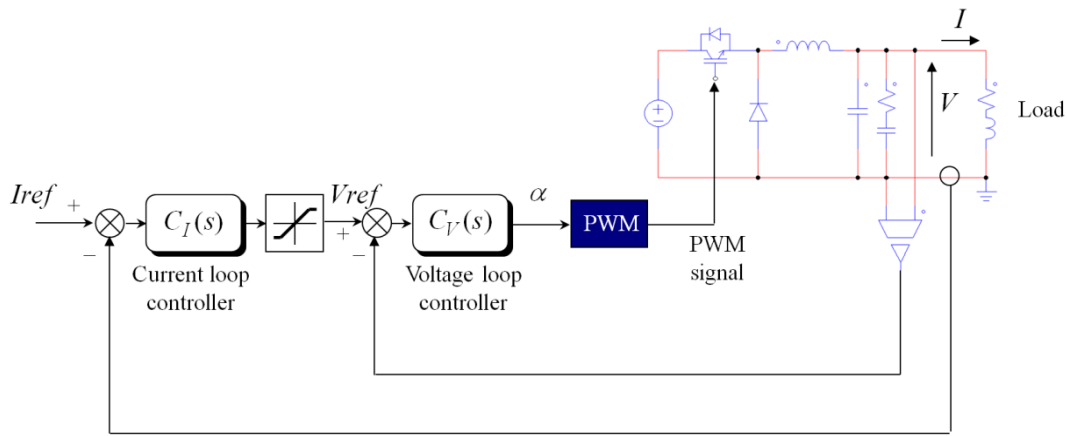

**Fig. 8:** Control of a Buck converter

Loop shaping is one of the most elementary methods used to design classical controllers such as PIDs. The controller is determined by manipulating (shaping) the open-loop frequency response, such that it meets the design specifications.

Let us assume the following specifications for the voltage loop:

– zero static error,
– dynamic precision, i.e., open-loop gain higher than 40 dB up to $w_0$,
– bandwidth $\{0, w_c\}$,
– phase margin $\Phi_M$ greater than 50°.

The plant model does not have a pole at $s = 0$ (integrator): it is necessary that the controller contains an integral action to cancel the static error. So a PI controller is first tried for $C_V(s)$. Its parameters are adjusted to meet the open-loop gain constraints, as shown on the Bode plot in Fig. 9.

However, with this setting the phase margin requirement cannot be reached. A derivative action is then added to correct the phase margin and meet the stability requirement.

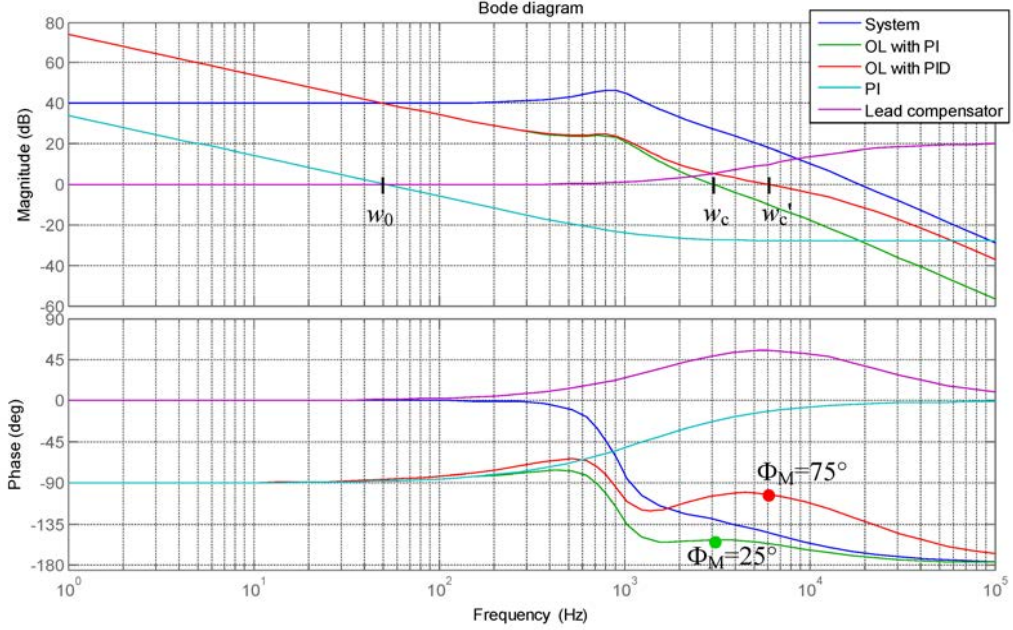

**Fig. 9:** PID tuning using the loop shaping method

Pole placement is used in the following to design a PI controller for the current loop. This method consists of placing the closed-loop poles at the desired positions by calculating the controller parameters such that this condition is fulfilled.

The open-loop transfer function is

$$OL_I(s) = C_I(s) \cdot CL_V(s) \cdot \frac{b_0}{1+a_1 \cdot s}, \qquad b_0 = 1/R_{\text{load}}, \quad a_1 = L_{\text{load}}/R_{\text{load}}, \tag{21}$$

where $CL_V(s)$ is the closed voltage loop transfer function.

If the current loop bandwidth is low compared to the voltage loop bandwidth, Eq. (21) can be simplified as follows:

$$OL_I(s) \approx k_p \cdot \left(1+\frac{k_i}{s}\right) \cdot \frac{b_0}{1+a_1 \cdot s}. \tag{22}$$

The system pole can be cancelled by setting $k_i = 1/a_1$. The closed-loop transfer function then shrinks down to a first-order transfer function:

$$CL_I(s) = \frac{1}{1+(a_1/k_p \cdot b_0) \cdot s}. \tag{23}$$

To make the closed-loop behave like a first-order system with a rise time equal to $t_r$, the closed-loop pole has to be placed at $-w_n$, where $w_n = 2.2/t_r$. This results in the following choice for the parameter $k_p$:

$$k_p = a_1 \cdot w_n / b_0. \tag{24}$$

The pole cancellation technique requires a good knowledge of the process: if $a_1$ is likely to vary ($L_{load} = f(I)$), this method can bring about poor results (especially when the pole is close to the origin). In that case, system pole cancellation should be avoided. The closed loop transfer function is then of second order. The PI parameters can be computed by identifying the coefficients of the closed-loop transfer function denominator with those of the polynomial $s^2/w_n^2 + (2 \cdot \zeta/w_n) \cdot s + 1$, where $w_n$ is the desired cut-off frequency (which can be related via Eqs. (13) and (15) to rise time and settling time) and $\zeta$ is the damping ratio (which has to be chosen greater than 1 for aperiodic behaviour). We obtain

$$k_p = (a_1 \cdot 2 \cdot \zeta \cdot w_n - 1)/b_0, \qquad k_i = a_1 \cdot w_n^2/(k_p \cdot b_0). \tag{25}$$

This controller setting, however, gives rise to a zero $-k_i$ in the closed-loop transfer function, which may affect the transient response (see Section 2.3.6). A solution to this issue is to cancel this zero by filtering the reference, as shown in Fig. 10.

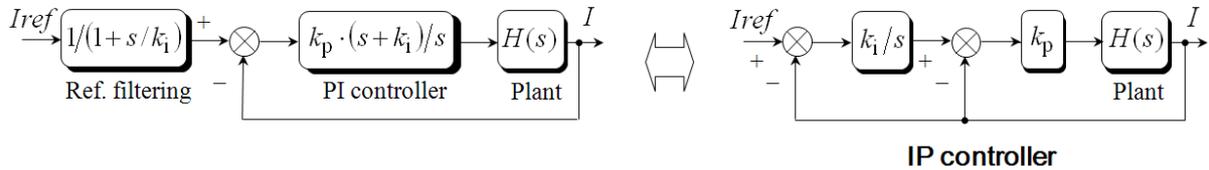

**Fig. 10:** Reference filtering

Another issue is the so-called *integral windup phenomenon*: when the output of either the current or the voltage controller saturates, the integral term of the controller increases and may become very large. At the end of the saturated mode of operation, a negative error is needed to remove the accumulated positive error, which may give large transients. Many anti-windup methods have been proposed [8–10]. A simple way to implement anti-windup is to switch off the integrator (by zeroing its input) in case of control output saturation. Other schemes, like the one represented in Fig. 11, enable the integral part of the controller to be dynamically saturated.

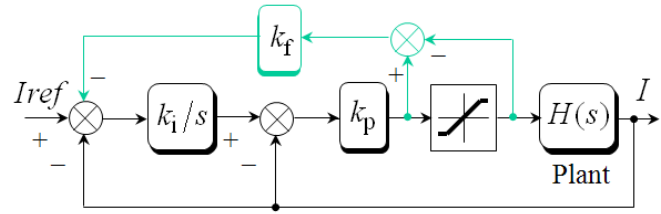

**Fig. 11:** Example of anti-windup solution

## 3    Discrete-time control

The development of digital technology over the past two decades along with increased performance demands have led to a growing use of digital control in power converters [11, 12]. The benefits of using digital over analogue control are numerous: performance enhancement (as digital control allows more complex regulation schemes), improved flexibility, system monitoring and archiving capabilities, better noise immunity, to name a few. One major issue is the delay introduced with a digital controller. Other issues, such as aliasing, quantization errors, or limit cycling, may also affect the system operation [13, 14].

## 3.1 Introducing the Z-transform

The Z-transform is the discrete-time counterpart of the Laplace transform. It is an essential tool for the analysis and design of discrete-time systems. Let us consider a sequence $x(k)$ of sampled signals, where $x(k)$ takes the value of the analogue signal $x(t)$ at the $k$th sampling instant (the sampling period $Ts$ is assumed constant). The Z-transform of this sequence is defined as

$$X(z) = \sum_{k=0}^{+\infty} x(k) \cdot z^{-k}, \tag{26}$$

where $z$ is a complex variable which is related to the Laplace operator by $z = e^{s \cdot Ts}$.

From the Laplace time-shifting property, we know that $e^{-s \cdot Ts}$ is the time delay by $Ts$ seconds. Therefore $z^{-1} = e^{-s \cdot Ts}$ corresponds to a unit sample period delay. The main properties of Z-transforms are listed below:

- linearity:

$$Z[\lambda \cdot x(k) + \mu \cdot y(k)] = \lambda \cdot X(z) + \mu \cdot Y(z), \tag{27}$$

- shifting property:

$$Z[x(k-n)] = z^{-n} \cdot X(z), \tag{28}$$

- convolution:

$$Z[x(k) * y(k)] = Z\left[\sum_{n=-\infty}^{n=+\infty} x(n) \cdot y(k-n)\right] = X(z) \cdot Y(z), \tag{29}$$

- final value:

$$\lim_{k \to \infty} x(k) = \lim_{z \to 1}(z-1) \cdot X(z). \tag{30}$$

Some examples of Z-transforms:

- discrete impulse: $X(z) = 1$,
- discrete step: $X(z) = \dfrac{1}{1 - z^{-1}}, \quad |z| > 1$,
- discrete ramp: $X(z) = \dfrac{z^{-1}}{\left(1 - z^{-1}\right)^2}, \quad |z| > 1$.

Tables of commonly encountered Z-transforms can be found in the literature (see the references).

## 3.2 Relation between Laplace transforms and Z-transforms

Let $x(t)$ be a causal continuous-time signal whose Laplace transform is $X(s)$, and let $x(k)$ be the discrete-time signal obtained by sampling $x(t)$ at a uniform rate $1/Ts$.

### 3.2.1 Case of signals having only simple poles

If $X(s)$ has only simple poles, it takes the following form:

$$X(s) = \sum_{i=1}^{N} \frac{A_i}{s - s_i}. \tag{31}$$

The expression of the corresponding Z-transform is

$$X(z) = \sum_{i=1}^{N} \frac{A_i}{1 - e^{s_i \cdot Ts} \cdot z^{-1}}. \qquad (32)$$

Thus, a pole $s_i = \sigma_i + j \cdot w_i$ in $X(s)$ gives rise to a pole $z_i = e^{s_i \cdot Ts} = e^{\sigma_i \cdot Ts} \cdot e^{j \cdot w_i \cdot Ts}$ in $X(z)$.

This relation enables one to establish the correspondence between the pole location in the s-domain and that in the z-domain.

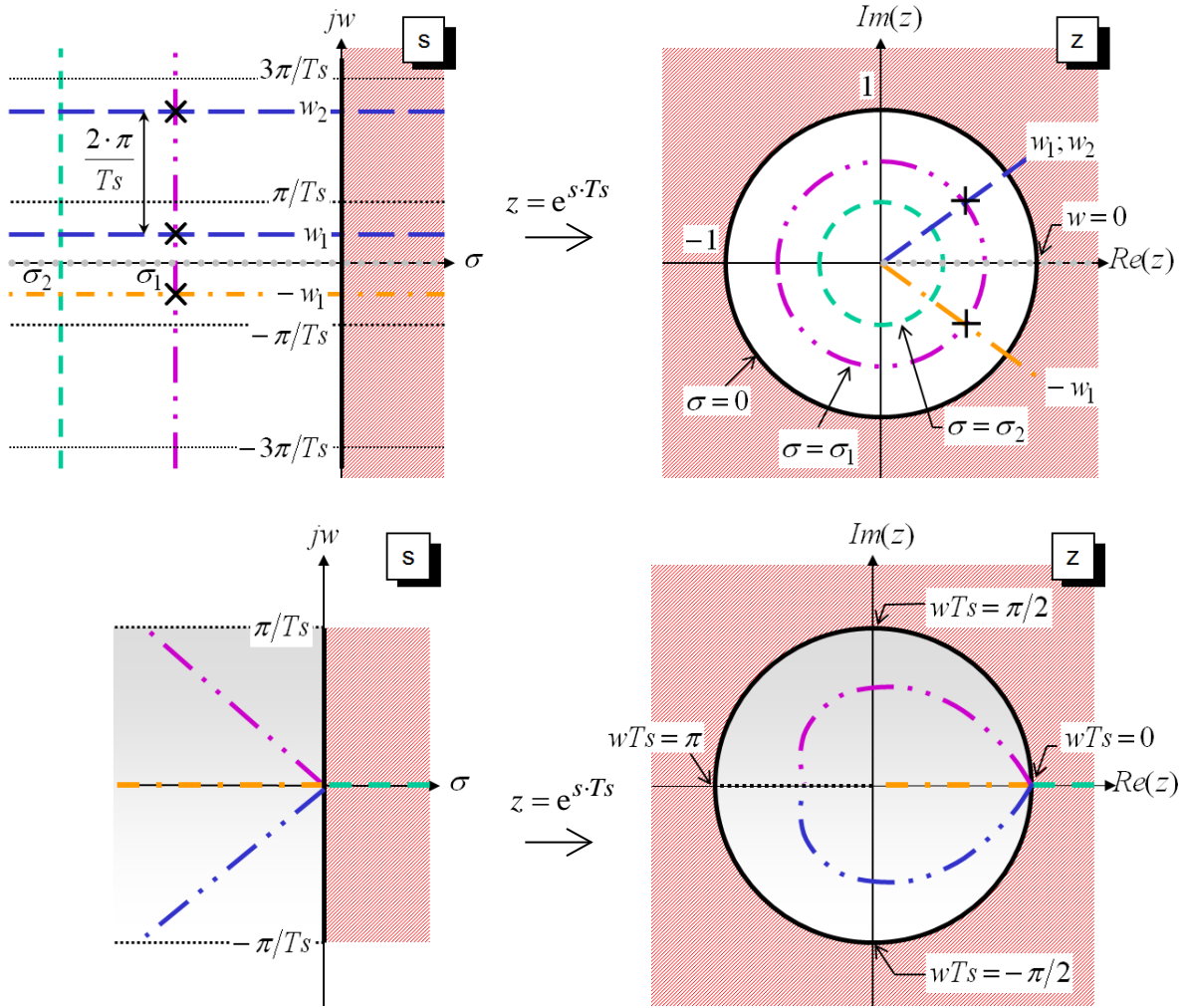

**Fig. 12:** Mapping of the *s*-plane to the *z*-plane

- The portion of the *s*-plane to the left of the imaginary axis maps to the inside of the unit circle in the *z*-plane, and the imaginary axis maps to the unit circle. So the stability region in the *s*-plane $\{\sigma_i < 0\}$ maps to $\{|z_i| < 1\}$.

- Lines of constant frequency in the *s*-plane become radial lines in the *z*-plane. The negative real axis in the *s*-plane maps to the interval 0 to 1 in the *z*-plane.

- Lines of constant real part map to circles centred at the origin.

- The lines of constant damping ratio in the *s*-plane become logarithmic spirals in the *z*-plane.

An important observation is that poles in the *s*-plane whose imaginary parts differ by a multiple of $2\cdot\pi/Ts$ belong to the same locations in the *z*-plane: the unit disc can only represent signals of frequency up to the half the sampling frequency (the Nyquist frequency). The condition which has to be satisfied so that two different poles in the *s*-plane do not correspond to the same point in the *z*-plane is
$$Ts < \pi \Big/ \max_i |w_i|. \qquad (33)$$

### 3.2.2 General case

In the general case, the relation between $X(s)$ and $X(z)$ is given by
$$X(z) = \sum_{s_i = \text{poles of } X(s)} \text{Residues}\left\{ X(s)\cdot \frac{1}{1-e^{s\cdot Ts}\cdot z^{-1}} \right\}_{s=s_i}. \qquad (34)$$

The residue at $s_j$, a pole of multiplicity $m$, can be calculated as follows:
$$\text{Residue}\left\{ X(s)\cdot \frac{1}{1-e^{s\cdot Ts}\cdot z^{-1}} \right\}_{s=s_j} = \frac{1}{(m-1)!}\cdot \lim_{s\to s_j} \frac{d^{m-1}}{ds^{m-1}}\left[ (s-s_j)^m \cdot X(s) \cdot \frac{1}{1-e^{s\cdot Ts}\cdot z^{-1}} \right]. \qquad (35)$$

## 3.3 Modelling of digitally controlled continuous-time systems

Let us focus on the physical process represented in Fig. 13, which is digitally controlled at the period *Ts*, and whose output is sampled at the same period. The structure of this hybrid system consists of a sampler (the analog-to-digital converter, or ADC), a digital controller, the continuous-time process, and a so-called hold-device (here assumed to be a digital-to-analog converter, or DAC). The sampler is preceded by an anti-aliasing filter (not represented in Fig. 13), which should preferably be taken into account for modelling.

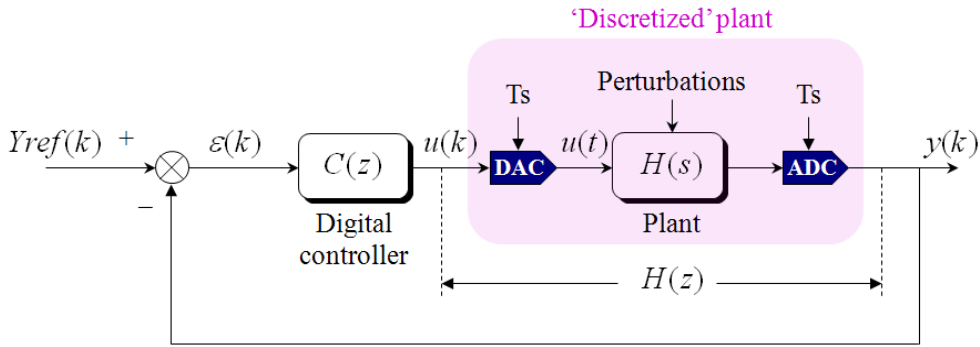

**Fig. 13:** Digitally controlled plant

The DAC converts $u(k)$ to $u(t)$ by holding each sample value for one sampling interval (zero-order hold operation). Hence it introduces a delay equal to $Ts/2$. The transfer function of a zero-order hold is
$$H_{\text{ZOH}}(s) = \frac{1-e^{-s\cdot Ts}}{s}. \qquad (36)$$

From an external point of view, the set {zero-order hold + plant + sampler} is a discrete system, whose transfer function can be derived from the plant transfer function according to
$$H(z) = (1-z^{-1})\cdot Z\left[\frac{H(s)}{s}\right]. \qquad (37)$$

It is interesting to note that if $H(s)$ has poles $s = s_i$, then the transfer function $H(z)$ of the 'discretized' system has poles $z = e^{s_i \cdot Ts}$. But the zeros are unrelated. Let us define

$$Y(z) = Z[y(k)] \; ; \; Yref(z) = Z[Yref(k)] \; ; \; E(z) = Z[\varepsilon(k)] \; ; \; U(z) = Z[u(k)].$$

The open-loop transfer function is then

$$OL(z) = \frac{Y(z)}{E(z)} = C(z) \cdot H(z) \tag{38}$$

and the closed-loop transfer function is

$$\frac{Y(z)}{Yref(z)} = \frac{C(z) \cdot H(z)}{1 + C(z) \cdot H(z)}. \tag{39}$$

The use of the operator $z^{-1}$ is often preferred as it refers to the past and not the future. The transfer function of the digital controller can then be written as

$$C(z) = \frac{U(z)}{E(z)} = \frac{b_0 + b_1 \cdot z^{-1} + \ldots + b_p \cdot z^{-p}}{1 + a_1 \cdot z^{-1} + \ldots + a_n \cdot z^{-n}}. \tag{40}$$

Hence,

$$(1 + a_1 \cdot z^{-1} + \ldots + a_n \cdot z^{-n}) \cdot U(z) = (b_0 + b_1 \cdot z^{-1} + \ldots + b_p \cdot z^{-p}) \cdot E(z). \tag{41}$$

Using the Z-transform shifting property, Eq. (28), the following difference equation is obtained, where the present output is dependent on the present input and the past inputs and outputs:

$$u(k) = b_0 \cdot \varepsilon(k) + b_1 \cdot \varepsilon(k-1) + \ldots + b_p \cdot \varepsilon(k-p) - a_1 \cdot u(k-1) - \ldots - a_n \cdot u(k-n). \tag{42}$$

This is the controller algorithm.

### 3.4 Analysis of closed-loop discrete systems

#### 3.4.1 *Stability and robustness of closed-loop discrete systems*

As seen in Section 3.2, a discrete system is stable if all the poles of its transfer function have a modulus strictly inferior to 1. So the closed-loop discrete system Eq. (39) is stable if all the roots of the characteristic equation $1 + C(z) \cdot H(z) = 0$ have a modulus strictly inferior to 1. The stability margins are defined as:

**Phase margin:**

$$\Phi_M = 180° + \arg\left[OL(e^{j \cdot w_{cr} \cdot Ts})\right], \tag{43}$$

where $w_{cr}$ is such that $\left|OL(e^{j \cdot w_{cr} \cdot Ts})\right| = 1$.

**Gain margin:** $\quad G_M = \dfrac{1}{\left|OL(e^{j \cdot w_\pi \cdot Ts})\right|}, \tag{44}$

where $w_\pi$ is such that

$$\arg\left[OL(e^{j \cdot w_\pi \cdot Ts})\right] = -180°.$$

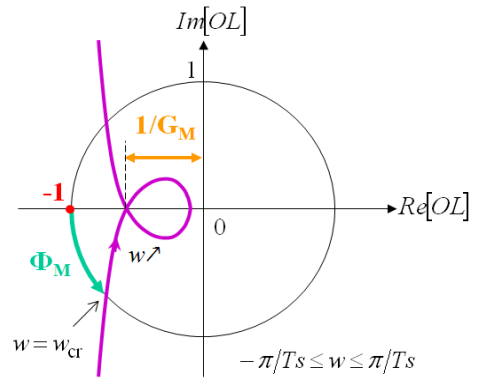

**Fig. 14:** Closed-loop discrete system stability

### 3.4.2 Influence of the poles on the system behaviour

The influence of the system poles on its transient behaviour is summarized in Fig. 15. Two interesting remarks are as follows:

− poles closer to the origin lead to a faster transient regime;
− the impulse response caused by poles having negative real part cycles back and forth between positive and negative deviations from the steady-state value.

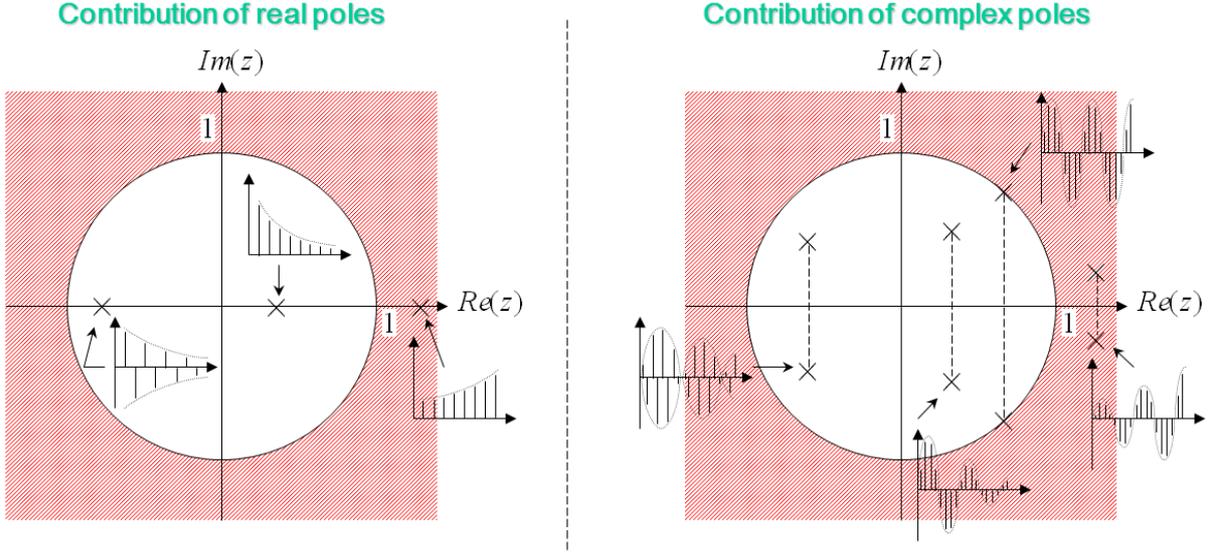

**Fig. 15:** Influence of the poles on the system transient behaviour

### 3.4.3 Particular case: second-order systems

As seen previously, a common controller design method consists in deriving the controller parameters from a pole placement, such that the dominant closed-loop dynamics is of first or second order. If a second-order system behaviour is chosen, this strategy consists in identifying the coefficients of the closed-loop transfer function denominator with those of the following reference polynomial:

$$Den_{CL_{des}}(z) = (1 - z_1 \cdot z^{-1}) \cdot (1 - \bar{z}_1 \cdot z^{-1}) \cdot P_{aux}(z), \tag{45}$$

where $z_1 = e^{-w_n \cdot Ts \cdot \left(\zeta - j \cdot \sqrt{1-\zeta^2}\right)}$ and $\bar{z}_1$ are the desired dominant poles, and $P_{aux}$ is a polynomial containing additional faster poles.

The choice of the dominant poles depends on the closed-loop behaviour specifications: the rise time and peak overshoot requirements impose the values of the cut-off frequency $w_n$ and the damping ratio $\zeta$ of the closed-loop system, according to Eqs. (13) and (14). The settling time requirement puts a constraint on the sampling period:

$$t_s = 4.6/\zeta \cdot w_n \implies |z_1| < e^{-4.6 \cdot Ts/t_s} \approx 0.01^{Ts/t_s}. \tag{46}$$

### 3.4.4 Precision of closed-loop discrete systems

The conclusions in the continuous-time domain remain valid: to achieve zero steady-state error, the open-loop transfer function must contain the internal model of the reference (for example, at least one integrators (pole at $z = 1$) is required for a step input). To reject perturbations of class $N$ in steady state, at least $N$ integrators are needed in the open-loop transfer function.

## 3.5 Discrete-time controller synthesis

There are two main strategies employed to design digital controllers:

- Emulation design, which consists in designing first a controller in the continuous-time domain. Afterwards, the controller is discretized such that the overall system preserves its characteristics.
- Direct discrete-time design, which consists in discretizing the plant at first using an appropriate hold device. Analysis and design can then be carried out in the discrete-time domain.

### 3.5.1 *Emulation design*

#### 3.5.1.1 *Principle of emulation design*

The first step is the continuous-time controller design. At this stage the sampling is ignored. The next step is the discretization of the controller by using an approximation (emulation) method enabling the approximation at best of the continuous-time controller. The usual approximation methods are the Euler method and the Tustin method (also called bilinear transformation), which consist in substituting *s* by Eqs. (47) and (48). The controller algorithm (difference equation) can then be derived.

Backward Euler transformation: $$s \to \frac{1}{T_s} \cdot (1 - z^{-1});\qquad(47)$$

Tustin transformation: $$s \to \frac{2}{T_s} \cdot \frac{1 - z^{-1}}{1 + z^{-1}}.\qquad(48)$$

As an example, let us discretize a PI controller using the Tustin transformation:

$$C(s) = k_p \cdot (1 + k_i/s);$$

$$C(z) = C(s)\big|_{s=\frac{2}{T_s}\cdot\frac{1-z^{-1}}{1+z^{-1}}} = k_p \cdot \left(\frac{1 + k_i \cdot T_s/2 + (-1 + k_i \cdot T_s/2)\cdot z^{-1}}{1 - z^{-1}}\right).$$

#### 3.5.1.2 *Comparison between Euler and Tustin approximation methods*

Both approximation methods preserve the stability, as shown in Fig. 16.

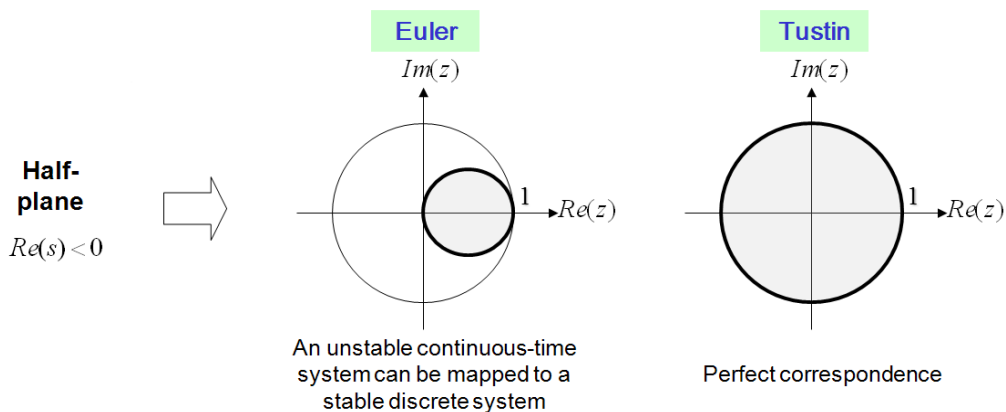

Fig. 16: Transformation of the stability region

Let us examine the image of a pole $s_0$ of a continuous-time system, obtained through these transformations:

- With the Euler transformation, the corresponding pole in the discrete-time domain is

$$z_0 = 1/(1 - s_0 \cdot Ts) = 1 + (s_0 \cdot Ts) + (s_0 \cdot Ts)^2 + (s_0 \cdot Ts)^3 + \ldots \quad (49)$$

- With the Tustin transformation, the corresponding pole is

$$z_0 = (1 + s_0 \cdot Ts/2)/(1 - s_0 \cdot Ts/2) = 1 + (s_0 \cdot Ts) + \frac{1}{2} \cdot (s_0 \cdot Ts)^2 + \frac{1}{4} \cdot (s_0 \cdot Ts)^3 + \ldots \quad (50)$$

Compared to the relation

$$z_0 = e^{s_0 \cdot Ts} = 1 + (s_0 \cdot Ts) + \frac{1}{2} \cdot (s_0 \cdot Ts)^2 + \frac{1}{6} \cdot (s_0 \cdot Ts)^3 + \ldots \quad (51)$$

Eqs. (49) and (50) constitute, respectively, a first-order and a second-order approximation. The higher the sampling frequency, the smaller the approximation error.

The poles (and the zeros) being not preserved, the system frequency response is changed. By way of example, the frequency responses of the derivative terms obtained through both transformations are shown in Fig. 17.

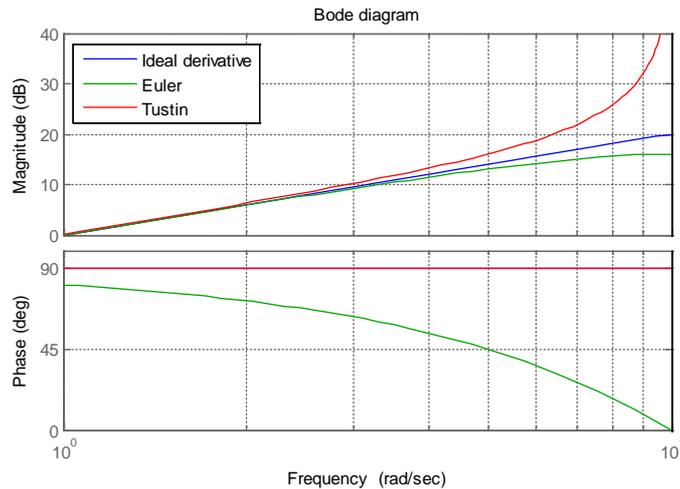

**Fig. 17:** Discretization of a pure derivative ($Ts = 0.3$)

It can be seen that there is an inherent filtering effect at high frequencies with the Euler transformation, whereas with the Tustin transformation the magnitude increases with the frequency, resulting in noise amplification at high frequencies. Thus the Euler transformation is more appropriate for the discretization of high-pass filters, and the Tustin approximation is more appropriate for the discretization of low-pass filters. To convert continuous-time controllers to discrete-time ones, the Tustin method is the approximation of choice, as it provides better accuracy.

To conclude, an emulation controller always suffers performance degradation as compared to its continuous-time counterpart. To make the set {ADC – digital controller – DAC} behave the same as the continuous-time controller, a very fast sampling frequency may be needed. Special attention should be paid to the impact on the loop phase margin of the control delay (ADC and DAC delays + delay due to the computation of control algorithm by the processor) and the anti-aliasing filter.

### 3.5.2 *Direct discrete-time design*

Working entirely with discrete systems, using plant discrete-time modelling, is another approach. This design method enables us to relax the constraint on the sampling period.

*3.5.2.1   Discussion on the choice of the sampling period*

A too small sampling period has the following consequences.

- Fast and expensive control hardware is required.
- Numerical issues may occur: the relation between poles in the s-domain and the z-domain is given by $z_i = e^{s_i \cdot Ts}$. So for $Ts \to 0$ we have $z_i \to 1 \; \forall s_i$. This may cause trouble when working in finite precision.
- The plant discretization may lead to the introduction of unstable zeros, which are difficult to compensate for.

On the contrary, a too large sampling period results in the following.

- Loss of information and aliasing, in the case of the violation of the Nyquist–Shannon sampling theorem.
- The regulation may not react readily enough to disturbances affecting the system.
- Plant discretization may yield poles having a negative real part: this is not desirable as the step response caused by such poles cycles back and forth between positive and negative deviations from the steady-state value. To prevent this, the following condition must be satisfied: $1/Ts > (2/\pi) \cdot \left| \mathrm{Im}(s_i) \right|$.

A suitable choice for the sampling period is

$$\frac{1}{Ts} \in \left[ 6 \cdot F_B^{CL} \; , \; 25 \cdot F_B^{CL} \right], \tag{52}$$

where $F_B^{CL}$ is the desired closed-loop bandwidth [2,14].

*3.5.2.2   RST-controller design*

The general structure of a discrete controller for linear single-input single-output systems is represented in Fig. 18.

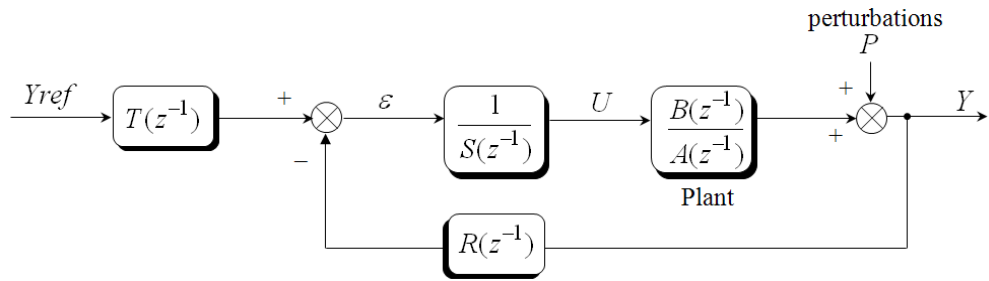

**Fig. 18:** RST-controller structure

This tri-branched structure is known as the RST structure: $R$, $S$, and $T$ are three $z^{-1}$ polynomials to be determined, usually by pole-zero placement. The control signal is calculated as

$$U(z) = \frac{T(z^{-1})}{S(z^{-1})} \cdot Yref(z) - \frac{R(z^{-1})}{S(z^{-1})} \cdot Y(z). \tag{53}$$

This is a combination of feedforward (first term) and feedback (second term), which can be tuned separately. The general approach with RST is to decouple the tracking behaviour (reference following) from the regulation behaviour (disturbance rejection); *R* and *S* give the regulation behaviour, and *T* gives the tracking behaviour [15–18].

The tracking transfer function is

$$\frac{Y}{Yref} = \frac{B \cdot T}{A \cdot S + B \cdot R} \tag{54}$$

and the regulation transfer function is

$$\frac{Y}{P} = \frac{A \cdot S}{A \cdot S + B \cdot R}. \tag{55}$$

Note that a PID controller is a special case of RST controller, where

$$Num_{PID}(z^{-1}) = R(z^{-1}) = r_0 + r_1 \cdot z^{-1} + r_2 \cdot z^{-2},$$

$$Den_{PID}(z^{-1}) = S(z^{-1}) = (1 - z^{-1}) \cdot (1 + s_1 \cdot z^{-1}),$$

$$T(z^{-1}) = R(z^{-1}).$$

The RST control law can be synthesized using the following design method:

1. Choice of the desired closed-loop transfer function:

$$CL_{des}(z^{-1}) = B_m(z^{-1})/A_m(z^{-1}) \tag{56}$$

If a second-order system behaviour is chosen, $A_m$ has the following form (see Section 3.4.3):

$$A_m(z^{-1}) = P_{dom}(z^{-1}) \cdot P_{aux}(z^{-1}) = (1 - z_1 \cdot z^{-1}) \cdot (1 - \bar{z}_1 \cdot z^{-1}) \cdot P_{aux}(z^{-1}), \tag{57}$$

where $z_1, \bar{z}_1 = e^{-\zeta \cdot w_n \cdot Ts} \cdot e^{\pm j \cdot w_n \cdot Ts \cdot \sqrt{1-\zeta^2}}$ are the dominant poles.

If a first-order system behaviour is desired, $P_{dom}$ is of the form

$$P_{dom}(z^{-1}) = (1 - a_1 \cdot z^{-1}), \tag{58}$$

where $a_1 = e^{-Ts/\tau}$, $\tau$ being the desired time constant.

2. Cancellation of stable poles and zeros of the plant transfer function:

Let us expand the plant transfer function:

$$H(z^{-1}) = \frac{B(z^{-1})}{A(z^{-1})} = \frac{B^-(z^{-1}) \cdot B^+(z^{-1})}{A^-(z^{-1}) \cdot A^+(z^{-1})} \tag{59}$$

- $A^-(z^{-1})$ contains the poles which will not be compensated, namely poles lying outside the unit circle (called unstable poles), poorly damped poles, and poles with negative real part.

- $B^-(z^{-1})$ contains the zeros which will not be compensated, namely zeros lying outside the unit circle (unstable zeros), the plant pure delay (of the form $z^{-d}$), poorly damped zeros, and zeros with negative real part.

$B^-(z^{-1})$ cannot be a factor of $A \cdot S + B \cdot R$ (the closed loop would be unstable). Thus $B_m(z^{-1})$ must be of the form

$$B_m = B^- \cdot B_{m_1}. \tag{60}$$

The pole–zero cancellation yields

$$R(z^{-1}) = A^+(z^{-1}) \cdot R_1(z^{-1}), \tag{61}$$

$$S(z^{-1}) = B^+(z^{-1}) \cdot S_1(z^{-1}), \tag{62}$$

$$T(z^{-1}) = A^+(z^{-1}) \cdot B_{m_1}(z^{-1}). \tag{63}$$

3. Perturbation rejection in steady state: an appropriate number of integral terms is introduced into the loop by means of the polynomial $S$:

$$S_1(z^{-1}) = (1 - z^{-1})^{n-l} \cdot S_2(z^{-1}), \tag{64}$$

where $n$ and $l$ are, respectively, the perturbation and plant classes.

4. Computation of $R$ and $S$: the polynomials $R_1(z^{-1})$ and $S_2(z^{-1})$, of the smallest possible degree, are determined by the resolution of the Diophantine equation:

$$A^-(z^{-1}) \cdot (1 - z^{-1})^{n-l} \cdot S_2(z^{-1}) + B^-(z^{-1}) \cdot R_1(z^{-1}) = A_m(z^{-1}) \tag{65}$$

Then $R$ and $S$ are derived from Eqs. (61), (62), and (64).

5. Computation of $T$ using Eq. (63). To ensure unity gain to the closed loop, the condition $T(1) = R(1)$ must be satisfied.

Let us return to the example used in Section 2.4.2.2—the RST controller for the outer current loop is designed in the following. The design specifications are:

– regulation behaviour—second-order system behaviour with a closed-loop bandwidth equal to $F_B^{CL}$;
– tracking behaviour—no overshoot.

The controller sampling period is chosen according to Eq. (52).

Let us assume that oversampling is used to improve resolution and reduce noise of the current measure, and that the cut-off frequency of the anti-aliasing filter is high compared to $F_B^{CL}$; hence there is no need to take it into account in the open-loop transfer function.

Let us also assume that the bandwidth of the inner voltage loop is high in comparison with the sampling frequency of the outer current loop: the digital model of the voltage source can be reduced to a simple unit gain.

Thus the plant transfer function shrinks down to

$$H(s) = e^{-tc \cdot s} \cdot \frac{b_0}{1 + a_1 \cdot s}, \text{ with } b_0 = 1/R_{load}, \quad a_1 = L_{load}/R_{load}, \tag{66}$$

where $tc = n \cdot Ts - \theta$ ($0 \leq \theta < Ts$) is the delay associated with conversion and computation times.

The discrete-time model can then be derived from Eq. (37):

$$H(z^{-1}) = (1-z^{-1}) \cdot Z\left[\frac{e^{-tc \cdot s}}{s} \cdot \frac{b_0}{1+a_1 \cdot s}\right] = b_0 \cdot z^{-n} \cdot (1-z^{-1}) \cdot Z\left[\frac{e^{\theta \cdot s}}{s} \cdot \frac{1}{1+a_1 \cdot s}\right].$$

Using Eqs. (34) and (35), we have

$$Z\left[\frac{e^{\theta \cdot s}}{s} \cdot \frac{1}{1+a_1 \cdot s}\right] = \lim_{s \to 0}\left[\frac{e^{\theta \cdot s}}{1+a_1 \cdot s} \cdot \frac{1}{1-e^{s \cdot Ts} \cdot z^{-1}}\right] + \lim_{s \to -1/a_1}\left[\frac{1}{a_1} \cdot \frac{e^{\theta \cdot s}}{s} \cdot \frac{1}{1-e^{s \cdot Ts} \cdot z^{-1}}\right]$$

$$= \frac{1}{1-z^{-1}} - \frac{e^{-\theta/a_1}}{1-e^{-Ts/a_1} \cdot z^{-1}}.$$

Finally,

$$H(z^{-1}) = \frac{B(z^{-1})}{A(z^{-1})} = b_0 \cdot z^{-n} \cdot \frac{1-e^{-\theta/a_1} + (e^{-\theta/a_1} - e^{-Ts/a_1}) \cdot z^{-1}}{1-e^{-Ts/a_1} \cdot z^{-1}}. \tag{67}$$

Let us assume that $Ts \ll a_1$ (sampling period small compared to load time constant) and $tc \ll Ts$. Equation (67) becomes

$$H(z^{-1}) = \frac{B(z^{-1})}{A(z^{-1})} \approx \frac{b_0}{a_1} \cdot z^{-1} \cdot \frac{Ts}{1-z^{-1}} = \frac{k_1 \cdot z^{-1}}{1-z^{-1}}. \tag{68}$$

An integrator is placed into the polynomial $S(z^{-1})$, i.e., $S(z^{-1}) = (1-z^{-1}) \cdot S_1(z^{-1})$. The polynomials $S_1$ and $R$ are then calculated by solving the following Diophantine equation:

$$(1-z^{-1})^2 \cdot S_1(z^{-1}) + k_1 \cdot z^{-1} \cdot R(z^{-1}) = A_m(z^{-1}) = (1-z_1 \cdot z^{-1}) \cdot (1-\bar{z}_1 \cdot z^{-1}), \tag{69}$$

where $z_1, \bar{z}_1 = e^{-\zeta \cdot w_n \cdot Ts} \cdot e^{\pm j \cdot w_n \cdot Ts \cdot \sqrt{1-\zeta^2}}$, $w_n = 2 \cdot \pi \cdot F_B^{CL}$, and $\zeta \geq 1$. We obtain

$$S_1(z^{-1}) = 1 \Rightarrow S(z^{-1}) = 1-z^{-1}, \tag{70}$$

$$R(z^{-1}) = r_0 + r_1 \cdot z^{-1}, \text{ with } r_0 = \frac{2-(z_1+\bar{z}_1)}{k_1}, \quad r_1 = \frac{z_1 \cdot \bar{z}_1 - 1}{k_1}. \tag{71}$$

If a simple gain is chosen for $T$, there is no undesirable zero in the tracking transfer function (no overshoot). To ensure unity gain to the closed loop, we must have

$$T(z^{-1}) = r_0 + r_1. \tag{72}$$

If a good tracking of the reference (with no lagging error) is required, the polynomial $T(z^{-1})$ can be fixed as follows:

$$T(z^{-1}) = A_m(z^{-1})/k_1 = (1-z_1 \cdot z^{-1}) \cdot (1-\bar{z}_1 \cdot z^{-1})/k_1. \tag{73}$$

Then the tracking transfer function becomes

$$\frac{Y(z)}{Yref(z)} = \frac{B(z^{-1}) \cdot T(z^{-1})}{A(z^{-1}) \cdot S(z^{-1}) + B(z^{-1}) \cdot R(z^{-1})} = \frac{k_1 \cdot z^{-1} \cdot T(z^{-1})}{A_m(z^{-1})} = z^{-1}, \qquad (74)$$

which means that the control loop replicates exactly the reference with one sampling step delay. This control technique, called dead-beat control, is very attractive since it can theoretically provide the fastest dynamic response for a digital implementation [18–20]. However, it requires an accurate modelling of the plant (polynomials *A* and *B*). In the case of model and parameter mismatches, significant deviations from the expected closed-loop performance can take place. Therefore, identification from experimental data may be necessary to refine the plant model. Moreover, if the plant parameters are likely to vary (e.g., magnet saturation), adaptive control with on-line tuning of the RST parameters may be required.

## 4   Conclusion

A brief review of regulation loops design in the continuous-time domains has first been made: the choice of the closed-loop performance has been discussed. PID controller synthesis using classical methods has been presented along with illustrating examples. Two different approaches to synthesize digital controllers have then been examined. The first one basically comprises translating an existing analogue controller (generally of PID type) into a digital one by using an approximation method. A good approximation of the continuous-time controller may imply the use of a high sampling frequency. The second method consists in synthesizing the digital controller by working with discrete systems exclusively. The sampling frequency is then chosen in accordance with the desired closed-loop bandwidth. This enables a better use of the potentialities of digital control, as the decrease of the sampling frequency favours the implementation of more complex control algorithms (RST type, requiring greater computational time), and the introduction of better digital signal filtering.